\documentclass[%
 aps,
 pra,%
 amsmath,amssymb,
reprint,%
longbibliography
]{revtex4-1}

\usepackage{graphicx}
\usepackage{dcolumn}
\usepackage{bm}
\usepackage[english]{babel}
\usepackage{lipsum}
\begin{document}


\title{A low-cost, helmet-based, non-invasive ventilator for COVID-19}

\author{Yasser Khan}
\email{yasser.khan@stanford.edu}
\affiliation{ 
Department of Chemical Engineering, Stanford University, 443 Via Ortega, Stanford, CA 94305‐4125, USA.}

\author{Hossain Mohammad Fahad}
\affiliation{ 
Serinus Labs, Inc. 2150 Shattuck Ave, Berkeley, CA, 94704-1370, USA.}

\author{Sifat Muin}
\affiliation{ 
Department of Civil and Environmental Engineering, University of California, Berkeley, CA 94720, USA}

\author{Karthik Gopalan}
\affiliation{ 
Department of Electrical Engineering and Computer Sciences, University of California, Berkeley, California 94720, USA.}

\date{\today}

\begin{abstract}
Coronavirus disease 2019 (COVID-19) has created an unprecedented need
for breathing assistance devices. Since the demand for commercial,
full-featured ventilators is far higher than the supply capacity, many
rapid-response ventilators are being developed for invasive mechanical
ventilation of patients. Most of these emergency ventilators utilize
mechanical squeezing of bag-valve-masks or Ambu-bags. These ``bag
squeezer'' designs are bulky and heavy, depends on many moving parts,
and difficulty to assemble and use. Also, invasive ventilation requires
intensive care unit support, which may be unavailable to a vast majority
of patients, especially in developing countries. In this work, we
present a low-cost (\textless{}\$200), portable (fits in an 8"x8"x4" box), non-invasive ventilator (NIV), designed to provide relief to
early-stage COVID-19 patients in low-resource settings. We used a
high-pressure blower fan for providing noninvasive positive-pressure
ventilation. Our design supports continuous positive airway pressure
(CPAP) and bilevel positive airway pressure (BiPAP) modes. A common
concern of using CPAP or BiPAP for treating COVID-19 patients is the
aerosolization of the severe acute respiratory syndrome coronavirus 2
(SARS-CoV-2). We used a helmet-based solution that contains the spread
of the virus. Our end-to-end solution is compact, low-cost
(\textless{}\$400 including the helmet, viral filters, and a valve), and
easy-to-use. Our NIV provides 0-20 cmH\textsubscript{2}O pressure with
flow rates of 60-180 Lmin\textsuperscript{-1}. We hope that our report
will encourage implementations and further studies on helmet-based NIV
for treating COVID-19 patients in low-resource settings.

\end{abstract}


\maketitle

The coronavirus disease 2019 (COVID-19) is a systemic disease that
primarily injures the respiratory system and may cause severe alveolar
damage and progressive respiratory failure \cite{Chan_2020,Huang_2020}. Initial
reports from China suggested approximately 15\% of individuals with
COVID-19 develop moderate to severe disease and require hospitalization
and oxygen support, with a further 5\% who require admission to an
intensive care unit (ICU) and supportive therapies including intubation
and mechanical ventilation \cite{Wu_2020}. But mortality rates were
found to be high among hospitalized COVID-19 patients requiring
intubation \cite{Richardson_2020}. On the other hand, early interventions
with oxygen therapies and breathing supports have shown to reduce the
need for intubation and complications in respiratory failure patients
while increasing survival rates \cite{Patel_2016}. In this scenario,
non-invasive ventilation becomes a viable measure for managing COVID-19
patients with acute respiratory distress syndrome (ARDS), early in the
disease.

\begin{figure*}
\includegraphics[width=18cm]{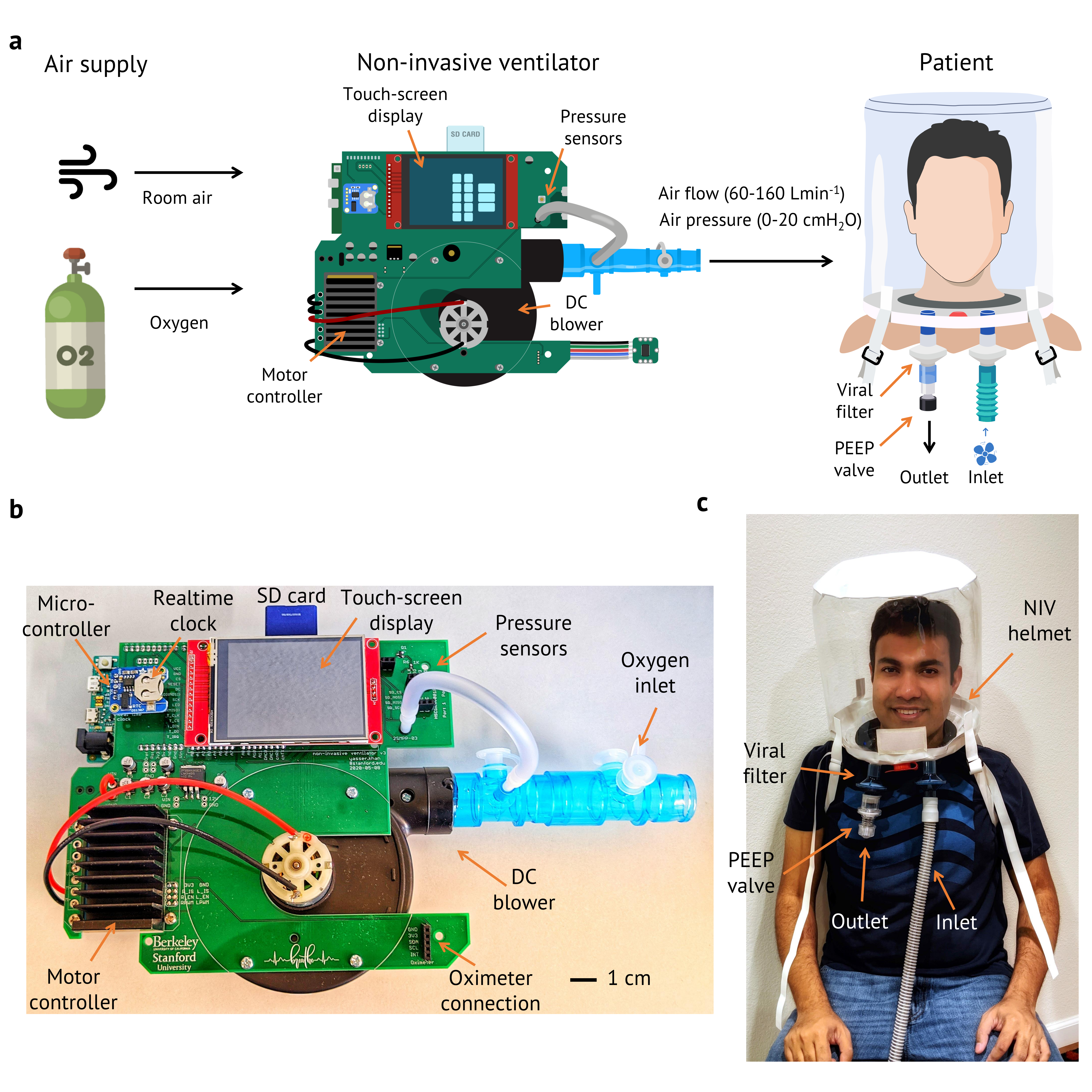}
\caption{\label{fig:1}\textbf{Overview of the helmet-based, non-invasive ventilator (NIV) for
		COVID-19}. (\textbf{a}) Schematic showing the helmet-based(NIV). Room
	air and 100\% oxygen are mixed according to the~fraction of inspired
	oxygen (FiO\textsubscript{2}) requirements. The NIV is controlled using a
	touch-screen display, where the user chooses the driving conditions of
	the direct current (DC) blower. A motor controller drives the DC blower.
	The whole system is operated using an Arduino Due microcontroller. Two
	pressure sensors monitor the pressure and flow of the NIV. The output of the blower connects to the inlet of the helmet. Two viral filters are used on the inlet and outlet sides to contain the severe acute respiratory syndrome
	coronavirus 2 (SARS-CoV-2). Besides, a positive end-expiratory pressure
	(PEEP) valve is used at the outlet. (\textbf{b}) Photograph of the NIV.
	The Arduino Due microcontroller is mounted from the back. Different
	parts for the NIV are shown - realtime clock, secure digital~ ~ ~(SD)
	card, touch-screen display, pressure sensors, DC blower, motor
	controller, and a port for connecting an oximeter. (\textbf{c})
	Photograph of a person wearing the helmet with the essential components
	for helmet-based NIV. The air out from the NIV connects to the inlet of
	the helmet through a viral filter. At the outlet side, a viral filter
	contains the spread of SARS-CoV-2, and a PEEP valve determines the
	PEEP setting of the whole system.}
\end{figure*}

Non-invasive ventilation refers to breathing support delivered through a
non-invasive method instead of invasive approaches, such as intubation
or tracheostomy. It provides positive air pressure to the patient to
recruit collapsed alveoli, improving blood oxygen levels, as well as
reducing carbon dioxide levels. In developing countries such as
Bangladesh, the use of full-featured mechanical ventilators (MVs) is
hindered by their high cost ($\approx$\$15,000 per unit),
limited availability, and requirement of associated ICU beds and expert
staffing for operation. Currently, there are about 1267 MVs available
for use in Bangladesh \cite{shortage}. But the demand for ventilators imposed by
COVID-19 at its peak is projected to be much greater.

To address the insurmountable need for ventilators, many rapid-response
ventilators are being developed for invasive mechanical ventilation. Most of these emergency ventilators use mechanical squeezing
of bag-valve-masks or Ambu-bags -- originally developed at Massachusetts Institute of Technology (MIT) in 2010 \cite{Al2010}. These ``bag squeezers'' do not need compressed
air or O\textsubscript{2}, however, they are bulky and heavy, depends on
many moving parts, and difficulty to assemble and use. These practical
challenges are huge impediments toward using bag-squeezer-ventilators
for supporting COVID-19 patients over many days. An alternate approach
is to use conventional NIVs, which are significantly less expensive with
no increased mortality associated with the usage \cite{covid-19,Garmendia_2020}. The availability of non-invasive respiratory support will be valuable for many patients or as a temporary bridge.

One of the major concerns of using NIVs such as continuous positive airway
pressure (CPAP) and bilevel positive airway pressure (BiPAP) for
COVID-19 patients is the risk of the aerosolization of the severe acute
respiratory syndrome coronavirus 2 (SARS-CoV-2), and transmission from
the patients to the health care workers (HCWs) \cite{Arulkumaran_2020}. The
main reason for generating aerosols using CPAP and BiPAP is the high
pressure (\textgreater{}10 cmH\textsubscript{2}O ) used in these
devices. SARS-CoV-2 can travel further and stay longer in the air when
transmission occurs through aerosols \cite{van_Doremalen_2020}. Therefore, in
order to minimize exposure to the virus, proper provisions are
necessary.

It is recommended to operate CPAP and BiPAP NIVs for treating COVID-19 patients in negative pressure rooms, which ensures SARS-CoV-2 is contained within
the room, and the HCWs are advised to take adequate protective measures.
Hence, NIV usage faces further constraints. Merging helmet-based
solution with noninvasive positive-pressure ventilation (NIPPV) is an
elegant solution that addresses the aerosolization of SARS-CoV-2 by
containing the virus inside the helmet.

In this work, we report a low-cost, portable, plug and play, NIV which
can provide breathing support to COVID-19 patients in low resource
settings. The device is an end-to-end solution with a helmet interface.
The helmet, which is equipped with virus filters, minimizes virus
exposure to HCWs while providing support to patients in need of positive
pressure ventilation. The NIV is portable (fits in an 8"x8"x4" box)
and has an easy-to-use touch-screen interface. Using a high-pressure
blower fan, we can provide CPAP and BiPAP modes. Our NIV provides 0-20
cmH\textsubscript{2}O pressure with flow rates of 60-160
Lmin\textsuperscript{-1}. Furthermore, our whole system is
\textless{}\$400, which includes the NIV, a helmet, viral filters, and
a PEEP valve, which is 40 times less expensive than an MV and 5 times
less expensive than a conventional NIV making it a fitting solution for
low-to-medium income countries like Bangladesh.

\section{\label{sec:results}Results}

\subsection*{Helmet-based, non-invasive ventilator for COVID-19} 

Ventilators provide precisely measured mix of room air and oxygen to
the patients with pressure and flow controls. Typically compressed air
and oxygen sources are connected to the ventilator unit. Depending on
the fraction of inspired oxygen (FiO\textsubscript{2}) requirements, room air and
oxygen are mixed in different ratios. The bag squeezer ventilators
provide air flow by the mechanical squeezing of Ambu-bags. Our NIV uses a direct current (DC) blower for providing adequate air pressure and flow. Figure 1 provides
an overview of the fully integrated helmet-based NIV system used in this
work. The setup is shown schematically in Figure 1a. Room air and oxygen
are mixed in the NIV. A port is provided to the user to connect to an oxygen
source, while room air is automatically used by the blower. Typically, an oxygen cylinder with a flow upto 15 Lmin\textsuperscript{-1} is
connected, while it is possible to connect two oxygen cylinders with a
flow up to 30 Lmin\textsuperscript{-1} to increase FiO\textsubscript{2}. The air outlet of the NIV connects
to the inlet of the helmet. Two viral filters are used at the inlet and outlet connections to contain the SARS-CoV-2. Furthermore, a positive end-expiratory pressure (PEEP)
valve is used at the outlet.

The integrated control electronics module is shown in Figure 1b. The
control electronics for the NIV incorporate commercial off-the-shelf
(COTS) components for control, sensing, and actuation and can operate in
either CPAP or BiPAP mode. An Arduino Due micro-controller is at the
heart of the system which controls a 12 V DC air blower to deliver
pressurized air to the patient helmet. The blower used here is a 50W air
pump that can generate a maximum static pressure of 28 cmH\textsubscript{2}O and a
maximum flow rate of 240 Lmin\textsuperscript{-1} at a current draw of 4 A. The blower speed/pressure is controlled by Arduino enabled pulse width
modulation (PWM) at 1 kHz and a high power motor driver capable of
providing at least 4 A of DC. A 2.8" touch-screen provides a simple and
intuitive user interface to the system, where clinically important
parameters such as pressure, inspiratory:expiratory
(I:E) ratio, breaths per minute (BPM) can be set and monitored.
This touch-screen also comes integrated with a secure digital (SD) card reader, providing
the option for data collection during patient treatment. The blower pressure
is monitored using a board mount COTS sensor with a -50 kPa to 50 kPa
operating range and a sensitivity of 1 kPa$\cdot$mV\textsuperscript{-1}. Additionally, this
pressure sensor is also used to detect leaks and tubing disconnects. A
board mount differential pressure sensor with a operating range of -6 kPa to 6 kPa is used to measure the compressed air flow rate. An additional function of this differential
pressure sensor is in the automatic triggering of inspiration and
expiration to enable patient-system synchronization in the BiPAP operating
mode. To indicate potential system or user faults such as pressure leaks
or accidental tubing disconnections, a magnetic DC buzzer is used to
generate an audible alarm. Figure 1c shows a person hooked up to the
helmet-based NIV. The filters and the PEEP valve are also shown in the
photograph.

\begin{figure*}
\includegraphics[width=17.4cm]{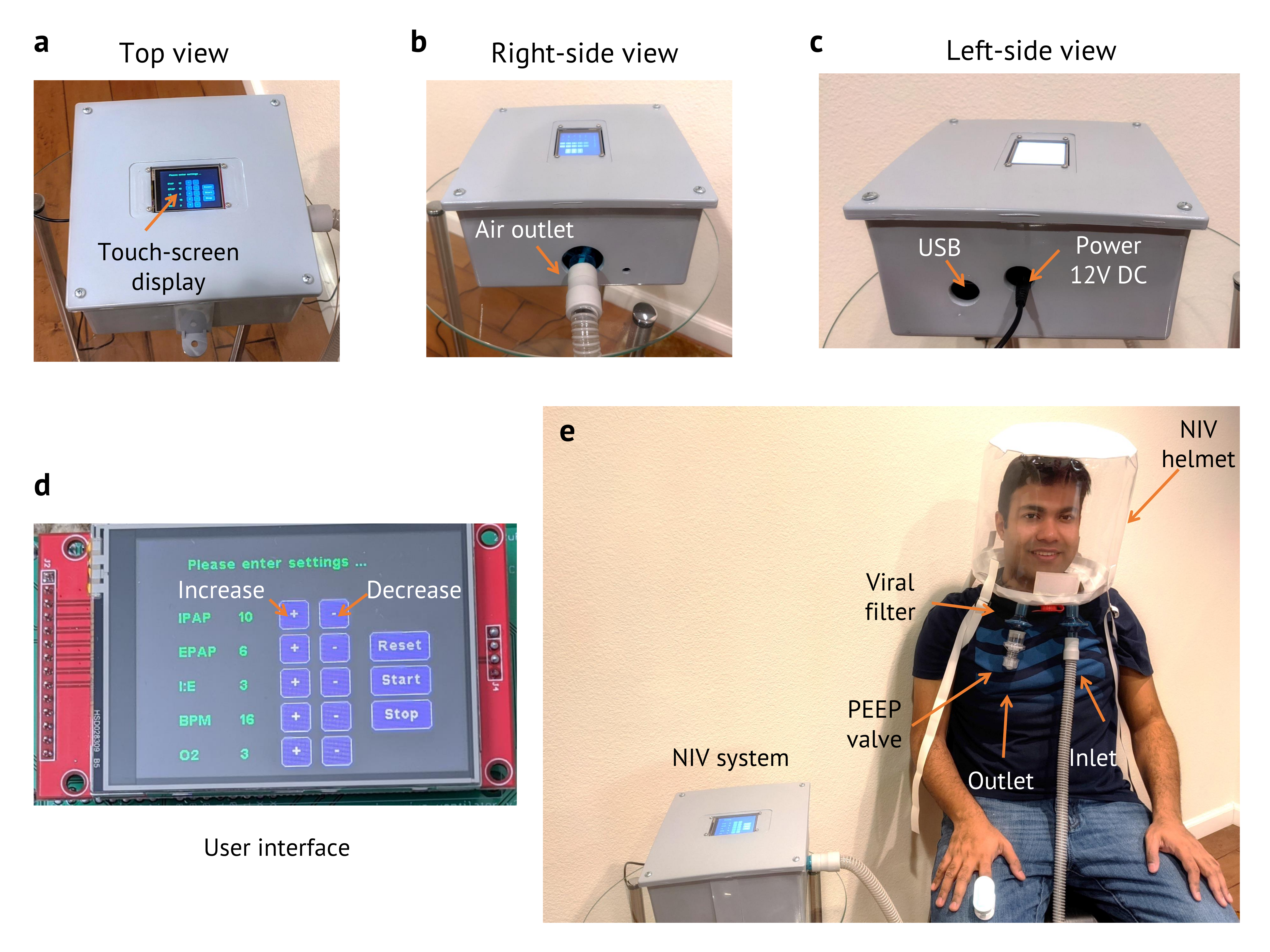}
\caption{\label{fig:oled} \textbf{Casing, user interface, and connection of the helmet-based NIV}.
	(\textbf{a}) Top view of the NIV casing showing the touch-screen display
	and the buzzer. (\textbf{b}) Right-side view of the NIV casing. Air outlet
	from the NIV, the oxygen inlet, and the oximeter connection are shown.
	(\textbf{c}) The left-side view of the NIV casing. Universal serial bus (USB) connection, DC
	power port, and the power switch are shown. (\textbf{d}) The user
	interface of the NIV. The user can set IPAP, EPAP, I:E ratio, and BPM.
	(\textbf{e}) Photograph of the complete setup showing the NIV housed
	inside an 8"x8"x4" polyvinyl chloride (PVC) enclosure, and the different components of air connections to the helmet.}
\end{figure*}

\begin{figure*}
\includegraphics[width=16.9cm]{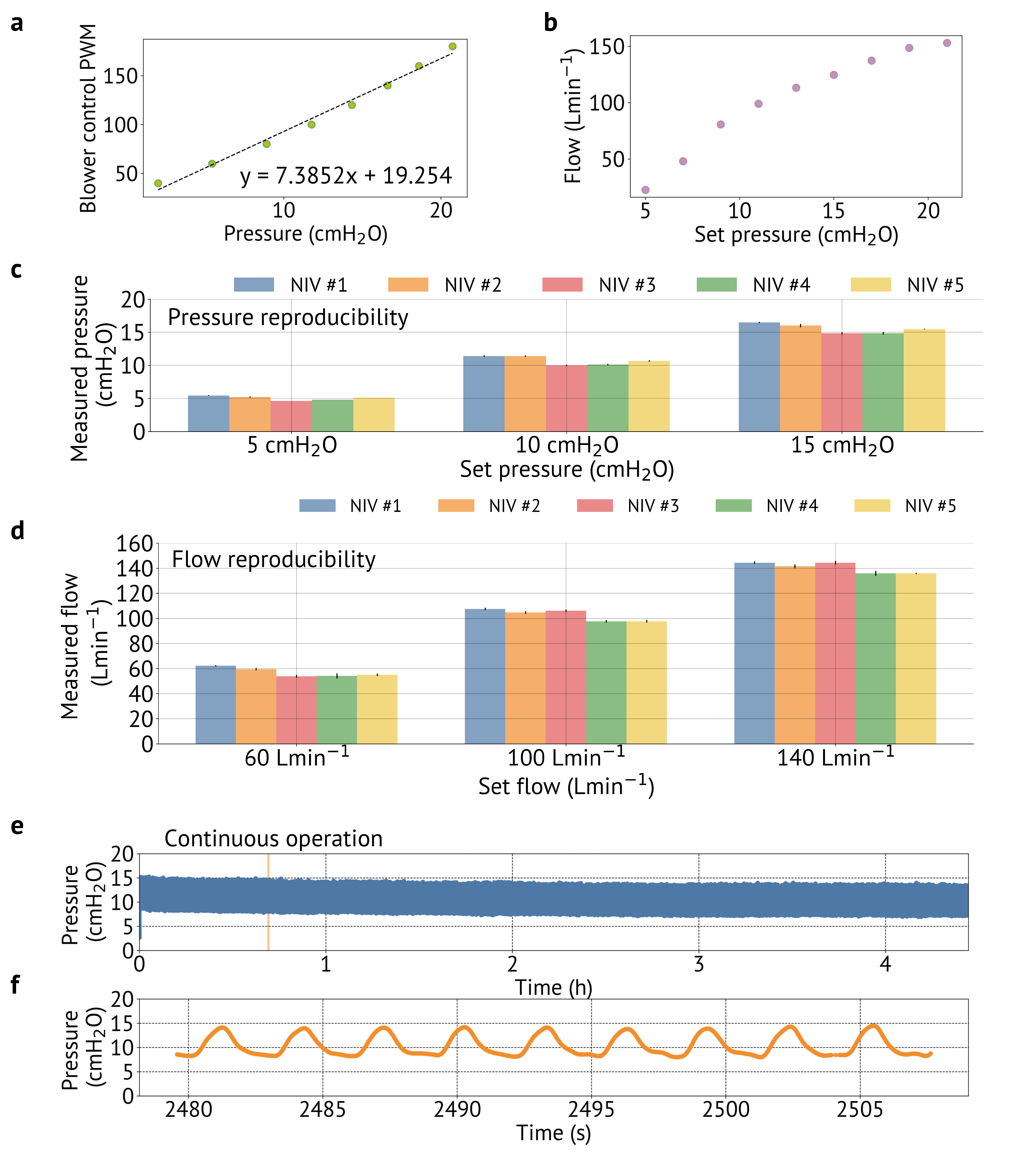}
\caption{\label{fig:opd} \textbf{Pressure, flow, and continuous operation characterization}. (\textbf{a})
	Pressure vs. blower control PWM plot. The desired pressure from the
	blower is obtained by selecting the corresponding control PWM signal.
	(\textbf{b}) Flow vs. pressure plot in an open-ended setting. The flow
	changes based on the set pressure, PEEP settings of the helmet, and the
	respiration of the patient. (\textbf{c}) Pressure reproducibility of 5
	different NIVs. 5, 10, and 15 cmH\textsubscript{2}O pressure were set and measured for each NIV. We observed 1 cmH\textsubscript{2}O variation without
	feedback control. (\textbf{d}) Flow reproducibility of 5 different
	NIVs. 60, 100, and 140 Lmin\textsuperscript{-1} flow rate were set and measured for each NIV. We observed 10 Lmin\textsuperscript{-1} variation without feedback
	control. (\textbf{e}) To test the continuous operation of the NIVs, we continuously operated
	one NIV at IPAP 15 cmH\textsubscript{2}O and EPAP 8 cmH\textsubscript{2}O for over 4 hours. (\textbf{f})
	Zoomed-in in data for 25 s do not show any visible fluctuation from the set
	pressure values.}
\end{figure*}
\begin{figure*}
\includegraphics[width=16cm]{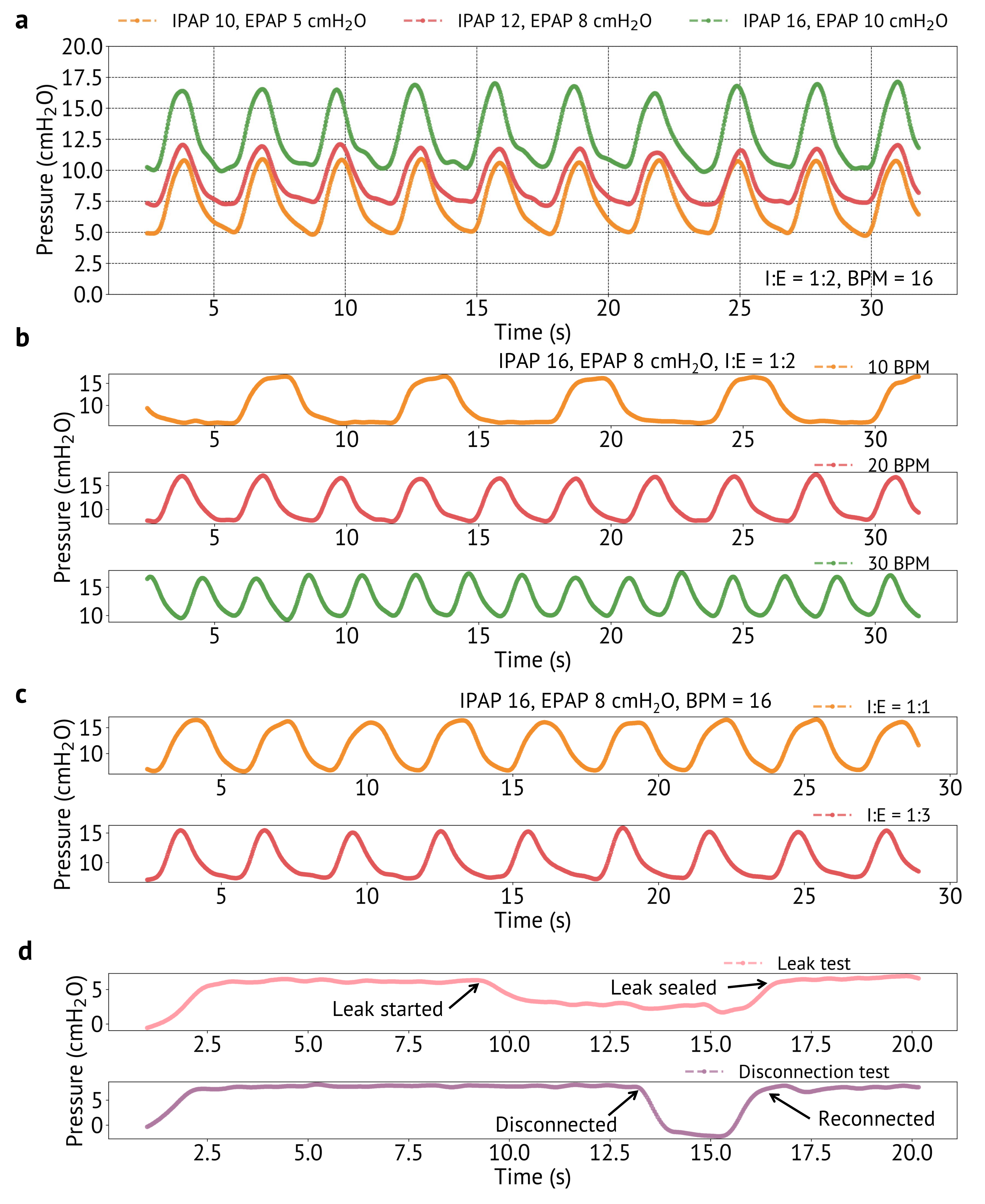}
\caption{\label{fig:comparison} \textbf{Pressure profile characterization}. (\textbf{a}) Different pressure profiles of (i)	IPAP 10 and EPAP 5 cmH\textsubscript{2}O, (ii) IPAP 12 and EPAP 8 cmH\textsubscript{2}O, and (iii) IPAP 16 and EPAP 10 cmH\textsubscript{2}O were tested. All profiles were within 0.5 cmH\textsubscript{2}O error margin. (\textbf{b}) BPM of 10, 20, and 30 were simulated to test different respiration rates of patients. (\textbf{c}) I:E ratios of 1:1 to 1:3 were simulated. (\textbf{d}) Leak and disconnection tests preformed using the helmet setup, a CPAP pressure of 12 cmH\textsubscript{2}O was set, keeping PEEP valve at 5 cmH\textsubscript{2}O. A leak was gradually generated which resulted in a pressure drop from 5 to 3 cmH\textsubscript{2}O. Similarly, disconnection was simulated at the inlet hose, which resulted in a pressure drop to 0 cmH\textsubscript{2}O is a matter of a second.}
\end{figure*}

The entire NIV system is integrated onto a custom printed circuit board
(PCB) and housed inside an 8"x8"x4" polyvinyl chloride (PVC) enclosure, as depicted in
Figure 2a-c. The touch-screen display and the buzzer are placed on the top
of the casing, while Air outlet from the NIV, the oxygen inlet, and the
oximeter connection are located on the right-side of the casing. One
oxygen cylinder with a flow rate up to 15 Lmin\textsuperscript{-1} can be connected to the
oxygen inlet, while it is possible to connect two oxygen cylinders with
a flow rate up to 30 Lmin\textsuperscript{-1}. A wired custom oximeter can be used with the NIV. On the left-side, universal serial bus (USB) port for programming the Arduino, DC power jack (12 V), and the power switch are located.

User inputs are taken with the 2.8" touch-screen as shown in Figure 2d.
The user can set inspiratory positive airway pressure (IPAP), expiratory
positive airway pressure (EPAP), I:E ratio, and BPM. IPAP and EPAP pressures can be set from 5-20 cmH\textsubscript{2}O, I:E
ratio of 1:1, 1:2, and 1:3 are supported. BPM can be adjusted from 10 to
30. Finally, the complete system setup with the NIV and the helmet is
displayed in Figure 2e.

\subsection*{Pressure and flow characterization}

Pressure is the most important parameter in CPAP / BiPAP-based NIVs.
Also, helmet-based ventilation require high-flow, to reduce CO\textsubscript{2}
rebreathing. A minimum flow of 60 Lmin\textsuperscript{-1} is required for helmet-based NIVs. We control the pressure setting of our NIV using PWM control signal, which is generated by the micro-controller and sent to the motor driver. Figure 3a shows the pressure vs. blower control PWM plot. While the blower reaches up to 24 cmH\textsubscript{2}O at PWM settings of 255, the pressure values are non-linear beyond PWM settings of 180. Therefore, we fixed the maximum pressure at 20 cmH\textsubscript{2}O for our NIV. The open-ended flow and pressure plot
is provided in Figure 3b. It is possible to reach up to 160 Lmin\textsuperscript{-1} flow
with the control settings, while the absolute maximum of the blower is
240 Lmin\textsuperscript{-1}. The flow depends on the set pressure, PEEP settings of
the helmet, and the respiration of the patient. Since the flow of 60 Lmin\textsuperscript{-1} occurs at a pressure of 7 cmH\textsubscript{2}O, we recommend a pressure difference of 7 cmH\textsubscript{2}O for operating the helmet-based NIV.

The pressure and flow reproducibility testing were done with 5 different
NIVs (Figure 3c and d). Pressures of 5, 10, and 15 cmH\textsubscript{2}O, and flows of 60,
100, and 140 Lmin\textsuperscript{-1} were set for the 5 NIVs. In the case of pressure testing, we
observed 1 cmH\textsubscript{2}O variation without feedback control. As for the flow testing, we
observed 10 Lmin\textsuperscript{-1} variation without feedback control. These
measurements were done using a one-time calibration shown in Figure 3a
and b. All the NIVs were operated using the same calibration curve.
Using individual calibration for each NIV will further improve the
variability from device to device.

Ventilators are operated for many hours to many days without
interruption. Therefore, it is essential to perform continuous operation testing of
the NIVs. We continuously operated one NIV at IPAP 15 cmH\textsubscript{2}O and EPAP 8 cmH\textsubscript{2}O for over 4 hours (Figure 3e), we did not observe any visible fluctuations from the set pressure values (Figure 3f). We are currently
doing lifetime testing of the device, which we will report in
the next version of the preprint.

\begin{figure*}
\includegraphics[width=18.5cm]{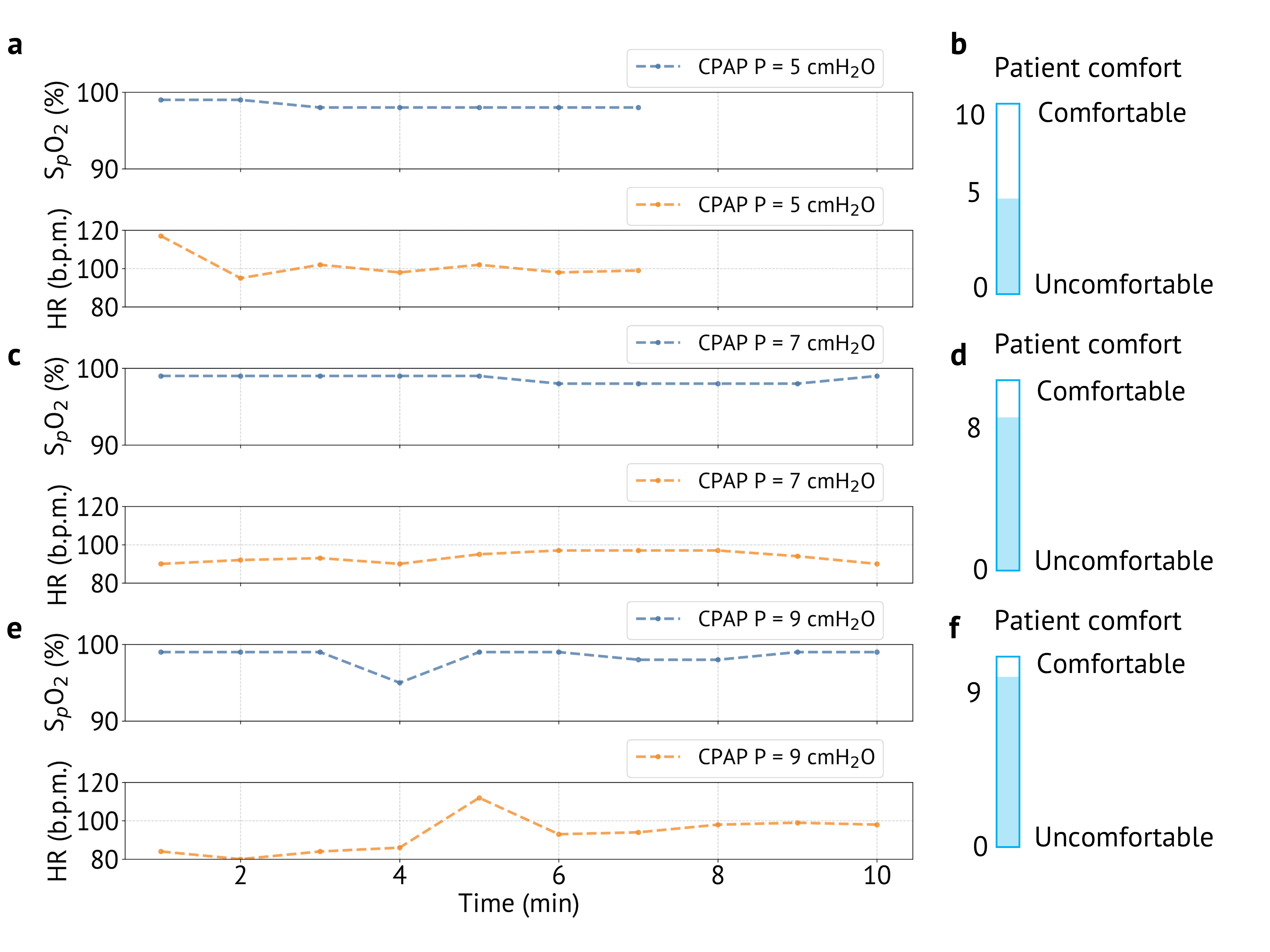}
\caption{\label{fig:4}\textbf{Human volunteer test results}. (\textbf{a}) Pulse oxygenation (SpO\textsubscript{2}) and heart rate
(HR) of the volunteer during the helmet-based NIV tests. CPAP pressure of
5 cmH\textsubscript{2}O was used in this experiment. (\textbf{b}) Comfort during the experiment reported on a 0-10 comfort scale. The volunteer reported comfort level of 5. (\textbf{c}) CPAP pressure of 7 cmH\textsubscript{2}O was used in this experiment. (\textbf{d}) The volunteer reported comfort level of 8. (\textbf{e}) CPAP pressure of 9 cmH\textsubscript{2}O was used in this experiment. (\textbf{f}) The volunteer reported comfort level of 9.}
\end{figure*}

\subsection*{CPAP and BiPAP modes for NIV}

Both CPAP and BiPAP modes can be used for non-invasive ventilation
\cite{Jaoude_2016}. In the CPAP mode, only one pressure is used. While
BiPAP switches between two different pressure settings. During the
inspiratory phase, higher pressure is used, and during the expiratory
phase, lower pressure is used. To characterize the performance of the
NIV, we operated the device in different IPAP, EPAP, I:E ratio, and BPM
settings as shown in Figure 4a-c. Different pressure profiles of (i)
IPAP 10 and EPAP 5 cmH\textsubscript{2}O, (ii) IPAP 12 and EPAP 8 cmH\textsubscript{2}O, and (iii) IPAP
16 and EPAP 10 cmH\textsubscript{2}O were tested with the device to simulate patients with varying levels of pressure needs. All the measurement were within
0.5 cmH\textsubscript{2}O error margin (Figure 4a). The respiration rate of patients can vary from 10 to 30 BPM, we simulated BPM of 10, 20, and 30, as shown in Figure 4b. We also tested varying I:E rations - 1:1 to 1:3, as shown in
Figure 4c. These characterization shows that our device can be operated
in BiPAP mode if the need arises.

One of the biggest safety features of NIVs is leak and disconnection
detection. To detect leaks, we used the on-board pressure sensor. Using the
helmet setup, a CPAP pressure of 12 cmH\textsubscript{2}O was set, keeping PEEP valve at 5 cmH\textsubscript{2}O. We then gradually created a leak as shown in Figure 4d, top panel. The pressure dropped from 5 to 3 cmH\textsubscript{2}O, which was detectable by the pressure sensor. This drop in pressure is used to trigger the leak alarm. Using the same setup we simulated disconnection of the inlet hose, the pressure goes to 0 cmH\textsubscript{2}O in a matter of a second (Figure 4d, bottom panel), this pressure signal is also used to trigger the alarm.

\subsection*{Initial human testing results}

\begin{figure*}
	\includegraphics[width=16cm]{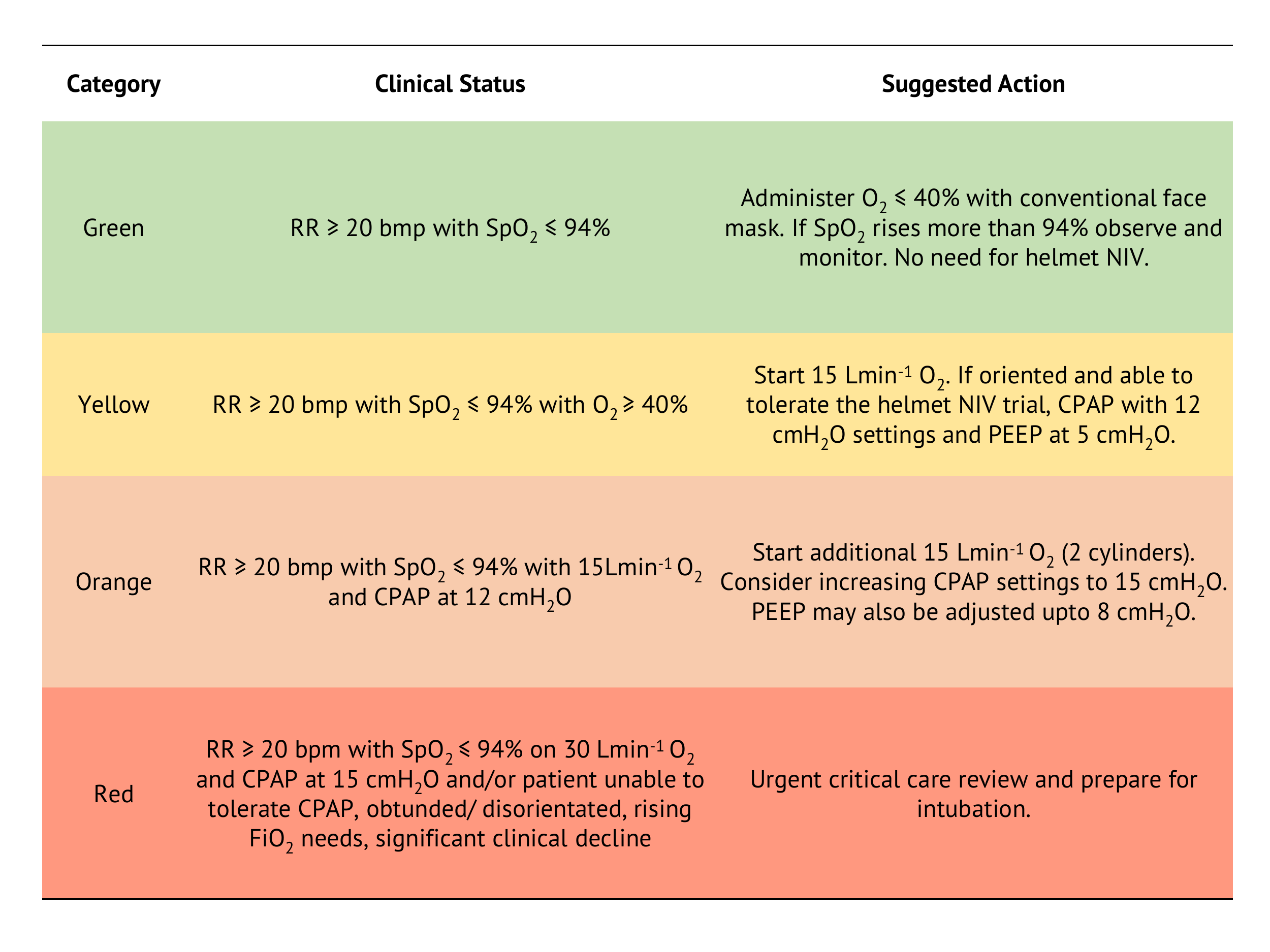}
	\caption{\label{fig:4}Adult escalation plan following initial assessment and treatment for COVID-19 patients in hospital. These are recommend based on the United Kingdom National Health Service (NHS) guidelines. We are actively working with doctors to update the protocol for Bangladesh considering the limited supply of oxygen cylinders with the maximum flow rate of 15 Lmin\textsuperscript{-1}.}
\end{figure*}

The efficacy of the NIV was tested by a human volunteer testing using the complete setup -- NIV with a helmet, viral filters, and a PEEP valve. These sets of experiments provided insight into the pressure and flow settings required for helmet-based NIV. We used CPAP pressure of 5, 7, and 9 cmH\textsubscript{2}O to test the low and medium flow regimes of the NIV (Figure 5). The volunteer wore a pulse oximeter for continuous monitoring of pulse oxygenation (SpO\textsubscript{2}) and heart rate (HR). The PEEP valve was set to 5 cmH\textsubscript{2}O. In the low flow regime, \textless{}30 Lmin\textsuperscript{-1}, helmet-based ventilation is uncomfortable as indicated by the volunteer -- rated 5 on a 0-10 comfort scale (Figure 5b). We also observed fogging of the helmet, which indicated inadequate flow. The pulse oximeter was used to monitor the SpO\textsubscript{2} level of the volunteer. SpO\textsubscript{2} went down slightly from 99\% to 97\% during the 7 min of testing. Then, low to moderate pressure settings of 7 and 9 cmH\textsubscript{2}O were used. The volunteer reported much higher
comfort levels (Figure 5c-e). Also, we did not observe fogging of the
helmet during the tests, which is a potential indicator for CO\textsubscript{2} re-breathing and insufficient flow. SpO\textsubscript{2} levels fluctuated between 99\% and 97\%, without any evident trend. These experiments verify that a minimum flow of 60Lmin\textsuperscript{-1} is required for helmet-based ventilation.

\subsection*{Proposed helmet-based non-invasive ventilation administration
	protocol}

In this section, we provide guidance for administering the developed
helmet-based NIV. This guidance is designed as a tool to aid the use of the
device alongside clinical judgment. It should not be treated as a
prescriptive measure. Decisions relating to the escalation of
ventilatory support need to be made by experienced clinical
decision-makers. United Kingdom National Health Service (NHS) guidelines have been used to outline the protocol \cite{suspected}.

Among different breathing support CPAP is the preferred form of NIV
support in the management of the hypoxaemic COVID-19 patient. Its use
does not replace invasive mechanical ventilation, but early application
may provide a bridge to it. Following are the suggested use, device
settings and monitoring guidelines for helmet-based CPAP:

\begin{enumerate}
	\item
	Suggested initial settings are 12 cmH\textsubscript{2}O + 15 Lmin\textsuperscript{-1} oxygen with a PEEP
	valve set at 5 cmH\textsubscript{2}O.
	\item
	Aimed SpO\textsubscript{2} is 94\% to 96\% for patients with acute respiratory	failure.
	\item
	Once CPAP/NIV has begun, the patient should be reviewed over 30 min
	to detect failed response or further decline. If the patient responds,
	hourly monitoring must continue for a further six hours. The frequency
	of assessment can be reduced if the patient is stable.
	\item
	Monitoring should focus on the regular measurement of respiratory
	rate, work of breathing, oxygen saturation, and heart rate.
	\item
	Consider increasing CPAP support to 15 cmH\textsubscript{2}O + 30 Lmin\textsuperscript{-1} 100\% oxygen (2
	cylinders) if needed. PEEP valve can be adjust to 8 cmH\textsubscript{2}O.
	\item
	If the condition remains stable or is improving, continue CPAP/NIV
	with a regular assessment.
	\item
	A low threshold for intubation should be established by the clinicians
	which may include a rising oxygen requirement, consistently rapid
	declining SpO\textsubscript{2}, consistently or rapidly increasing respiratory rate,
	and increased work of breathing. This should trigger immediate
	assessment for intubation and mechanical ventilation if deemed
	appropriate.
\end{enumerate}

This guidance is summarized in Figure 6.
\subsection*{Discussion}

The available resources to address ventilation needs are drastically
different in developing countries. With growing evidence that
noninvasive positive pressure ventilation results in a significant
reduction in endotracheal intubation \cite{Antonelli_2007}, it is a
promising low-resource treatment plan for treating COVID-19 patients in
developing countries. To facilitate this treatment approach, we presented an end-to-end, low-cost, helmet-based ventilator. Our device moves away from the bulky, heavy, difficulty to assemble and use bag-squeezer-ventilators by using a compact design that is portable and easy-to-use. Also, we mitigate the aerosolization of SARS-CoV-2 with a helmet-based approach. We hope we presented a compelling case for our NIV for use in medium to low-to-medium income countries. Nonetheless, further studies and randomized clinical trials are required to evaluate the efficacy and the benefits of helmet-based NIV in treating COVID-19. 

\subsection*{Disclaimer}

We released this version of the preprint due to the extraordinary
circumstances of COVID-19. The developed hardware and the results of
this study have not been medically approved. However, the system
engineering efforts and validation results indicate the potential
for relief to early-stage COVID-19 patients. We intend to
introduce helmet-based, low-cost, non-invasive ventilation to healthcare
professionals working under low-resource settings. Our hardware should
NOT be used for invasive mechanical ventilation. Furthermore, we are
improving the software to enhance the hardware usage of the current
platform. The next version of the preprint will report the on-going
enhancements.\\
A provisional patent application has been filed based on
the technology described in this work.

\subsection*{Acknowledgements}

This project was made possible by the generous support of the
Bangladeshi students and alumni of UC Berkeley and Stanford University.
We would like to thank Aurika Savickaite for providing us a helmet for
testing, Dr. Maurizio Cereda, Prof. Md Robed Amin, Dr. Tarik Reza, Dr. Tanveer Ahmad, and Dr. Nakib Shah Alam for helpful technical discussion, Prof. Taufiq Hasan, Faisal Huda, Sheikh Waheed Baksh, Dr. Raisul Islam, Dr. Shegufta
Mishket, Yasser M. Tahid Khan, Dr. Yaseer M Tareq Khan, Eusha Abdullah
Mashfi, Minhaz Khan, and Nicholas Vitale for helping us throughout the project. We would also like to thank Prof. Ana Claudia Arias and Prof. Miki Lustig
for allowing us lab-access for the fabrication of the casing. We also
thank Bay Area Circuits for providing us PCBs in an expedited manner.

\bibliographystyle{naturemag}
\bibliography{niv}

\end{document}